**Quantum fluctuations, quanta of electromagnetic interaction, quantum electronic bound states**

V.N. Murzin[1] and L.Yu. Shchurova[1]

[1]Lebedev Physical Institute, Russian Academy of Sciences, Leninskii Prospect 53, Moscow, Russia 119991

**Abstract**

Based on the concepts of the quantum field theory of virtual photons as quanta of electromagnetic interaction, we discuss the physical content of the phenomena underlying the principle of quantum uncertainties. We consider the features of the uncertainty relations and the properties of the elementary particles (electrons, protons, etc.) under the conditions of the formation of quantum bound states at atomic and subatomic distances.

**Key words**: virtual photons, uncertainty relations, bound states of an electron and a proton, electron-proton interaction at subatomic distances.

# 1 Introduction

The concept of a "quantum particle" characterized by dual wave-particle behavior and probabilistic description, is at the center of initial posits and postulates of quantum theory, formulated at the beginning of the 20th century. Although the equations of quantum mechanics themselves and the fact that they work perfectly were not questioned by the physicists, already in the late 1920s a discussion arose about the interpretation of the mathematical formalism of quantum mechanics. A group of physicists led by N. Bohr believed that the state of a quantum object does not exist objectively and independently of the measurement, and the probabilistic description of physical phenomena offered by quantum mechanics is complete. A. Einstein denied the essential role of observations in the quantum process and considered quantum mechanics to be incomplete (in the form in which it was formulated), since it does not provide a satisfactory explanation of the nature of quantum fluctuations and the probabilistic nature of quantum phenomena [1, 2]. Even now, many decades later, there is no consensus on the main issues related to the interpretation of quantum theory. V. Ginzburg, in his article [3] on problems of physics at the turn of the 20th-21st centuries, highlighted "the three great problems in physics". One of these problems is understanding of the foundations of quantum theory. The current state of quantum theory, as well as some open questions in quantum mechanics, including problems of the probabilistic nature of quantum objects and the influence of measurements on quantum processes, were discussed in recent papers [4, 5, 6].

Relativistic quantum mechanics and quantum field theory (in particular, quantum electrodynamics, or QED) [7, 8] form the foundation of modern physical views on the nature of microparticles, and allow to distinguish some theoretical aspects that can be used for deeper understanding of basic quantum phenomena.

In a number of publications devoted to these issues [9-15], the QED concepts of virtual photons, i.e. quanta of electromagnetic interaction, are applied to describe the physics of quantum uncertainties. In our earlier papers [14, 15] we analyzed the physical essense of the uncertainty relations based on the concept of virtual photons and their role in the formation of quantum properties of particles. Our consideration [14, 15] was carried out under the applicability conditions of non-relativistic quantum mechanics, that is, in the approximation of low velocities of interacting particles $v \ll c$, at long interparticle distances (greater than the atomic sizes) and for relatively small values of uncertainties (small deviations of dynamic variables from their stationary values). Non-relativistic quantum mechanics is invariant under the Galilean transformations, which are the small-velocity limit of the Lorentz transformations, and therefore it is not dedicated to describe the properties of quantum systems at high energies. Moreover, non-relativistic Quantum mechanics, in principle, does not address the problems of the formation of interaction processes [16], which are important in relativistic quantum theories [8, 10].

In this paper we discuss the specific features of the interaction processes involving virtual photons in the systems with relatively low distances between interacting particles shorter than the electron Compton length. We consider peculiarities of the uncertainty relations that can be expected from the point of view of the virtual photon concept under the condition of strong electromagnetic interaction of particles at subatomic distances. We particularly analyze the formation of quantum-stable bound states in a system of electrically charged particles, the role of virtual photons in the formation of such states, and the properties of bound states of particles such as electron and a proton.

The paper is organized as follows. In Section 2, using the concept of virtual photons, we discuss the quantum uncertainty, the Heisenberg uncertainty relations, and the underlying physics. In Section 3, we apply the concept of virtual photons to consider the features of the interaction processes and formation of quantum-stable bound states in a system of electrically charged particles (electrons, protons, etc.) at atomic and subatomic distances. We show that at subatomic distances, the Heisenberg's uncertainty relations change: they become less constraining and allow smaller values for the momentum uncertainty. When applied to the problem of motion of an electron in the Coulomb field of a proton, this leads to the possibility of a relativistic solution with nuclear scales of energies and distances. We compare the parameters of this relativistic solution with the characteristics of neutrons found in literature.

**2. Quanta of electromagnetic interaction and quantum uncertainties**

In QED, interactions of charged particles are described in terms of emission and absorption of virtual photons, i.e. of the quanta of electromagnetic interaction [7-9, 17]. The specific features of virtual photons, which distinguish them from the properties of real photons, are directly related to quantum uncertainty relations [18, 19]. It is well known that the laws of conservation of energy and momentum prohibit the emission and absorption of real photons by a particle (for example, an electron) at rest or moving with a constant speed. However, the Heisenberg uncertainty relations,

$$\Delta p_e \Delta r_e \geq \hbar, \qquad (1)$$

$$\Delta \varepsilon_e \Delta t_e \geq \hbar, \qquad (2)$$

allow for a momentum spread $\Delta p_e$ and energy spread $\Delta \varepsilon_e$ of an electron, and the possibility of electrons to emit and absorb virtual photons with the energy $\varepsilon_{ph} = \hbar \omega_{ph}$ for a short period of time $\Delta t_e \approx \hbar / \Delta \varepsilon_e \approx \hbar / \hbar \omega_{ph}$.

Virtual photons, unlike real photons, exist only for a short time interval $\Delta t \approx (1/2\pi) T_{ph}$ not exceeding the virtual photon wave period $T_{ph}$. It means that virtual photons cannot propagate away from a particle emitting photons at distances longer than the photon wavelength. As a consequence, virtual photons never completely leave the emitting particle: they are localized in a small spatial region $\sim (c/\omega_{ph})^3$ that includes both interacting particles providing the interaction between them [10, 14-15]. These features are taken into account by introducing a nonzero inertial mass $m_{ph}$ into the dispersion relation of virtual photons,

$$\varepsilon_{ph} = \sqrt{m_{ph}^{\ 2} c^4 + p_{ph}^{\ 2} c^2} \ , \tag{3}$$

where $\varepsilon_{ph}$ and $p_{ph}$ are the energy and 3-momentum of virtual photons, correspondingly. The 4-momentum of virtual photons, $P_{ph}$, is given by

$$P_{ph}^{\ 2} = \varepsilon_{ph}^{\ 2} / c^2 - p_{ph}^{\ 2}. \tag{4}$$

Virtual photons are thus attributed to nonzero inertial mass, $m_{ph} \neq 0$, which increases with the momentum $P_{ph}$ as $m_{ph} = \sqrt{|P_{ph}|^2 / c^2}$ [18] and provides the longitudinal polarization component, which leads to the formation of the electrostatic potential. At large inter-particle distances, this electrostatic potential coincides with the Coulomb potential $\varphi_{coul}(r) = e/r$.

As follows from (3)-(4), at large inter-particle distances and weak electromagnetic interaction, virtual photons are characterized by small values of the inertial mass $m_{ph} \ll p_{ph}/c$ and are close in their properties to real photons. At short distances (with a significantly stronger interaction), $m_{ph}$ increases and becomes comparable to or even exceeds the value of $p_{ph}/c$. Thus, at small distances, virtual photons may be considered to be similar to ordinary particles [15]. These characteristic properties of virtual photons allow us to comment on the physical content of the Heisenberg uncertainty relations under the applicability conditions of non-relativistic quantum mechanics.

In the approximation of relatively weak electromagnetic interaction (i.e. for small deviations of dynamic variables, $\Delta \varepsilon_e \ll \varepsilon_e$, $\Delta p_e \ll p_e$, $\Delta v_e \ll v_e$, $\Delta x_e \ll x_e$, where $v_e$ is the average velocity of the particle), the energy uncertainty of a particle (electron) is due to the action of a virtual photon on it: $\Delta \varepsilon_e \approx \hbar \omega_{ph}$ [14, 15, 18, 19]. The uncertainty of the particle momentum $\Delta p_e$ cannot be found from the virtual photon momentum, because the momentum values of virtual photons, which provide the interaction of particles, are negligible at large distances $r \gg r_C$, where $r_C = \hbar / m_e c \approx 3.8 \cdot 10^{-11} cm$ is the Compton wavelength of an electron. As was shown in [14], the use of the expression $\Delta \varepsilon_e = v_e \Delta p_e$ in this case is justified in the same way as in Refs. [16, 20, 21]: the expression $\Delta \varepsilon_e = v_e \Delta p_e$ follows from $\varepsilon_e = \sqrt{m_e^2 c^4 + p_e^2 c^2}$ and is applicable both in non-relativistic and relativistic cases. Thus the particle momentum uncertainty is expressed as $\Delta p_e \approx \Delta \varepsilon_e / v_e \equiv \hbar \omega_{ph} / v_e$ [14]. Uncertainties in the time and coordinate of a particle, $\Delta t_e \approx 1/\omega_{ph}$ and $\Delta x_e \approx v_e \Delta t_e \equiv v_e / \omega_{ph}$, are determined by the duration of the virtual photon action on a particle and the average velocity of the particle $v_e$ (assuming $\Delta v_e \ll v_e$) [10, 21, 22].

As a result, the effect of a virtual photon on a particle leads to uncertainty relations $\Delta p_e \Delta r_e \geq \hbar$ and $\Delta \varepsilon_e \Delta t_e \geq \hbar$ (coinciding with equations (1) and (2)) that do not depend on the virtual photon frequency and, therefore, can be considered as universal [14, 15]. The physical nature of these regularities is directly related to the universal ability of virtual photons (regardless their frequencies) to carry and transfer the same value of action $S_{ph}$ equal to the Planck's constant: $S_{ph} = \hbar \omega_{ph} T_{ph} = h$ [14, 15, 23].

Usually, an ensemble of virtual photons of different frequencies takes part in each quantum process. However, since the products of uncertainties are independent of the virtual photon frequency $\omega_{ph}$, the relations (1,2) are applicable not only to a single photon, but also to the whole ensemble of virtual photons involved in a quantum process.

Note that in the framework of a similar approach, two more uncertainty relations can be considered, i.e. $\Delta p_e \Delta t_e \geq \hbar / v_e$ and $\Delta \varepsilon_e \Delta r_e \geq \hbar v_e$. These relations also do not depend on the frequency of virtual photons, while they are less universal because they depend on the average particle velocity, $v_e$. These uncertainty relations can be regarded as complementary to the Heisenberg's relations as discussed in [14].

In conclusion of this section, let us stress the following points. The original formalism of quantum mechanics of the beginning of the 20th century does not include the concept of virtual photons which was developed later in the context of QED. In this work, quantum uncertainties and uncertainty relations are examined from the point of view of the QED concept of virtual photons. The validity of this approach is confirmed by establishing the invariance of the Heisenberg uncertainty relation to the parameters of virtual photons. Quantum uncertainties which are a manifestation of the probabilistic properties of quantum phenomena, receive a natural explanation within this approach. Namely, quantum indeterminacy and the uncertainty relations viewed in this way do not require information about "measurements" and "observers" for their interpretation, since they are completely determined by the processes of inter-particle interaction mediated by virtual photons.

**3. Features of the formation of quantum-stable bound states in a system of electrically charged particles, such as a proton and an electron, at atomic and subatomic distances within the virtual photons concept**

The formation of quantum bound states between particles at atomic and subatomic distances is an interesting research problem in atomic physics and relativistic QED [7-10]. Protons and neutrons have a complex internal structure: inside each proton or neutron, there are three quarks bound by gluons. Protons and neutrons are of about the same size, their root-mean-square radius is $\approx 0.8 \cdot 10^{-13} cm$ [24]. The "physical" radius of the free electron is known to be less than $10^{-20} cm$ [24, 25]. In atomic physics with relatively large interparticle distances, the internal structure of protons and neutrons can be disregarded, and the problem of bound electronic states can be considered within the framework of the electrodynamic theory of motion of protons and neutrons as elementary (structureless) particles.

In the case of arbitrary distances, the formation of bound states in a system of charged particles is a rather complicated problem already within the framework of QED. At small interparticle distances, the description of the interaction processes requires taking into account higher orders of perturbation theory, including creation/annihilation of electron-positron pairs, and other effects. However, at distances exceeding $10^{-13} cm$, the situation is quite accessible for QED analysis. According to QED calculations in [14], at such distances in a system of charged particles such as a proton and an electron, the Coulomb potential formed by virtual photons is realized, and the contribution of higher terms of the expansion does not exceed a fraction of a percent. When analyzing stationary bound electronic states, the time dependence is not as important (as compared to scattering processes), and, therefore, the analysis can be carried out in three-dimensional space using 3-momenta and 3-coordinates. There are two basic equations that should be considered when describing the formation of bound states [26]; the first one

$$p_e r = \hbar \qquad (5)$$

(where $r$ is the distance between a proton and an electron) is the quantum requirement that, in the case of bound electronic state, the uncertainties of the electron momentum $\Delta p_e$ and the uncertainty of the radius of its orbit $\Delta r_e$ in the uncertainty relations (1) и (2) are of the order of the momentum and radius themselves: $\Delta p_e = p_e$ and $\Delta r_e = r$. Equation (5) can be rewritten in a more convenient form as

$$\hbar \omega_{ph} = \hbar \omega_e^{orb} = \hbar \omega_e = p_e v_e, \qquad (6)$$

where $\omega_e^{orb} = v_e / r$ is the frequency of the electron orbit, and $\omega_e$ is the wave frequency of the electron. Expression (6) is equivalent to the requirement that the length of the electron orbit in the ground bound state equals to de Broglie electron wavelength.

As the second equation, we use

$$e^2 / r = p_e v_e. \qquad (7)$$

similarly to the description of the Lamb shift effect in a hydrogen atom [10, 12-13] in terms of virtual photons. Equation (7) describes the electron motion on the orbit of radius $r$ in the Coulomb field of a proton ($p_e = m_e v_e / \sqrt{1-(v_e/c)^2}$ is the 3-momentum of an electron) [16]. The solution of equations (6) and (7) is the well-known bound state of a proton and an electron in the Bohr model of the hydrogen atom. It is characterized by the electron velocity $v_e / c = \alpha \approx 0.0073$ (where $\alpha = 1/137$ is the fine structure constant), the orbital radius $r_B = \hbar^2 / m_e e^2 \approx 0.5 \cdot 10^{-9} cm$, and the binding energy $\varepsilon_{cs} = \alpha^2 m_e c^2 / 2 \approx 13.6 eV$.

Equations (6) and (7) allow to find the characteristic parameters of virtual photons that form the Bohr bound state of an electron and a proton. Taking into account the relation $\Delta\varepsilon_e = v_e \Delta p_e$ (and given that $\Delta\varepsilon_e = \hbar\omega_{ph}$), we rewrite equation (6) in the form:

$$\hbar\omega_{ph} = \hbar\omega_e^{orb} = \hbar\omega_e = p_e v_e. \tag{8}$$

Equation (8) shows that there is a direct relation between electron frequency, $\omega_e^{orb}$ and $\omega_e$, and the frequency $\omega_{ph}$ of virtual photons mediating the bound state: $\omega_{ph} = \omega_e^{orb} = \omega_e$. It also follows from (8) and (7) that in the case of the hydrogen atom, the Bohr bound state is formed by a narrow ensemble of virtual photons with a pronounced maximum energy close to the potential energy of the electron in the Coulomb field of the proton, $\hbar\omega_{ph} = e^2/r$. The wavelength of virtual photons that form the Bohr bound state of an electron and a proton significantly exceeds the distance between the electron and the proton: $\lambda_{ph}^B \approx 2\pi c/\omega_{ph} = 2\pi c\hbar/(e^2/r_B) = (2\pi/\alpha)r_B \succ 500 r_B$.

### 3.1. Features of the electron–proton Coulomb interaction at distances comparable and less than the electron Compton wavelength

A special feature of formation of bound states in the system of a proton and an electron at subatomic distances comparable to the Compton wavelength of electron is due to recoil effects. As is known, in the QED description of the interaction of electrically charged particles mediated by virtual photons, it is sufficient to take into account only the additive contributions of the instantaneous Coulomb interaction and transverse waves (see [17], pp. 4, 127). [27] analyzed the formation of the Coulomb potential in a system of particles, such as a proton and an electron, through the emission and absorption of photons with a fixed low mass. In the case of the quantum-stable bound state considered here, a similar situation is realized. The bound state of each particle is formed by virtual photons with a certain energy and fixed inertial mass. At a distance $r$ between two particles, the potential mediated by such photons is written as [27]:

$$\varphi(r_{12}) = \frac{e_1 e_2}{4} \frac{e^{-\eta \cdot r}}{r}, \tag{9}$$

where $\eta = m_{ph} c/\hbar$ is the inverse Compton wavelength of a photon with the mass $m_{ph}$. It follows from (9) that the Coulomb type of potential at a given distance $r$ is formed by virtual photons regardless their frequency and the inertial mass only if this distance does not exceed the Compton wavelength of the virtual photon ($r \leq \hbar/(m_{ph}c)$). At distances $r \approx 10^{-13} cm$, this situation is realized if the inertial mass of the virtual photons does not exceed $300 m_e$.

Expression (9) was derived under the assumption that the transfer of a virtual photon momentum to the interacting particle, or recoil effects, can be neglected [27]. Indeed, in the case of weak Coulomb interactions mediated by low-energy virtual photons, their momenta are small compared to the momenta of particles with a finite rest mass. However, it is well known that the recoil effect is a specific feature of Compton scattering, in which real photons interact with an electron and transfer some energy due to inelastic scattering to a recoil electron. This changes the frequency of scattered photons, and also decreases the Compton scattering cross section at photon energies $\hbar\omega \geq m_e c^2$ [23]. The decreased cross-section manifests itself at photon energies $\hbar\omega \geq m_e c^2$ and is described by the Klein-Nishina formula $\Phi = \zeta(\gamma)\Phi_0$ [23] for the total cross scattering of the electromagnetic radiation by an electron, where

$$\zeta(\gamma) = \frac{3}{4}\left\{\frac{1+\gamma}{\gamma^3}\left[\frac{2\gamma(1+\gamma)}{1+2\gamma} - \ln(1+2\gamma)\right] + \frac{1}{2\gamma}\ln(1+2\gamma) - \frac{1+3\gamma}{(1+2\gamma)^2}\right\}. \tag{10}$$

Here, $\gamma = \hbar\omega/m_e c^2 \leq 1$ is the dimensionless photon energy, $\omega$ is the photon frequency. The Thomson scattering cross section $\Phi_0 = 8\pi r_e^2/3$ (where $r_e = e^2/(m_e c^2) \approx 2.8 \cdot 10^{-13}\,cm$ is the classical radius of the electron) was chosen as the unit for $\Phi$.

The decrease in the Compton scattering cross section (compared to the Thomson one) described by the function $\zeta(\gamma)$ is associated with the effects of momentum exchange between photons and an electron. A part of the energy, which is transferred by photons to an electron as a result of the recoil effect, is spent mainly on increasing the electron velocity. Therefore, this energy is excluded from the radiation scattered by the electron. Similar situation arises in the case of the interaction of virtual photons with an electron in the formation of the Coulomb interaction between particles. In this case, part of the energy transferred by virtual photons to an electron, as in the case of real photons, is spent on the movement of the particle as a result of the recoil effect, but this is not taken into account when calculating the potential (9). This means that for virtual photons with not too large masses, the Coulomb potential at small distances is formed mostly by more energetic virtual photons with an energy $\hbar\omega_{ph} \geq e^2/r$, and not with the energy of $\hbar\omega_{ph} = e^2/r$ (as in the case of a weak Coulomb interaction) [14, 15].

By taking this into consideration, in Eq. (6) we should introduce the coefficient $\beta(r) \leq 1$, which depends on $r$ and accounts for the recoil effect:

$$\beta(r)\hbar\omega_{ph} = \beta(r)\hbar\omega_e^{orb} = \beta(r)\hbar\omega_e = p_e v_e, \quad \beta(r) \leq 1. \tag{11}$$

Obviously, the same coefficient $\beta(r)$ should be introduced into the Heisenberg uncertainty relation for the 3-momentum, $\Delta p_e \Delta r_e \geq \hbar$,:

$$\Delta p_e \Delta r_e = \beta(r)\hbar, \qquad \beta(r) < 1. \tag{12}$$

The uncertainty relation (12) that takes into account the recoil effect, becomes less restrictive than the usual Heisenberg uncertainty relation and allows for lower values of the momentum and coordinate uncertainties [15]. Figure 1 shows the dependence $\beta(r/r_C)$ calculated on the basis of equations (7) and (11) and using the relation (10) in the case of interacting electron and the proton.

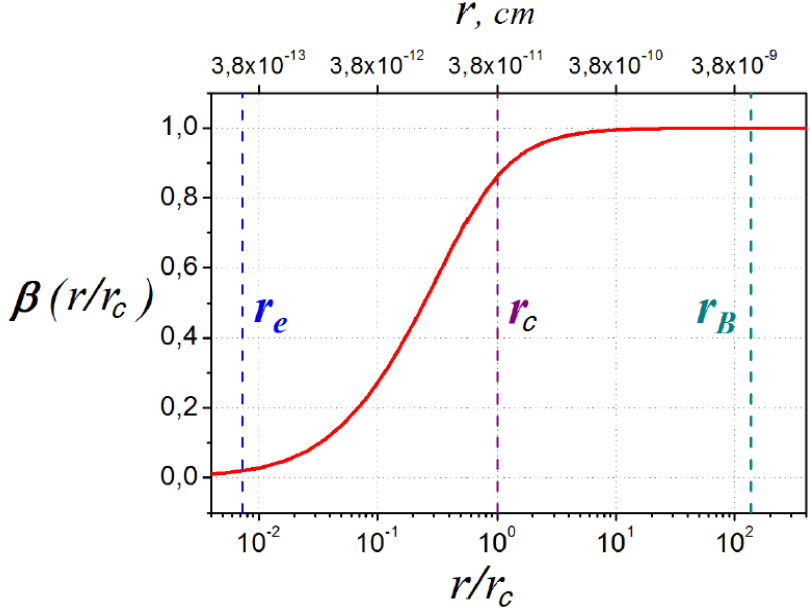

Figure 1. The calculated dependence $\beta(r/r_C)$; $r$ is the distance between the electron and the proton, $r_C = \hbar/(m_e c) \approx 3.8 \cdot 10^{-11} cm$ is the Compton wavelength of an electron, $r_e = e^2/(m_e c^2) \approx 2.8 \cdot 10^{-13} cm$ the classical electron radius, $r_B = \hbar^2/(m_e e^2) \approx 5.3 \cdot 10^{-9} cm$ is the Bohr radius.

## *3.2. Nonrelativistic and relativistic solutions to the problem of the formation of bound states of an electron in the Coulomb field of a proton at atomic and subatomic distances*

By taking into account the recoil effect, the uncertainty relation (11) and (12) leads to smaller values of the momentum and coordinate uncertainties than the usual Heisenberg relations. As a result, (11) and (12) provide the possibility of two bound-state solutions to the problem of electron motion in the Coulomb field of a proton.

Using the coefficient $\beta(r)$ calculated using expression (10), equations (7) and (11) lead to a **nonrelativistic** Bohr solution (at *β*=1) for the electron velocity $v_e/c = \alpha/\beta \approx 0.0073$ and the orbit radius $r = r_B = (c/v_e)^2 r_e \approx 5.3 \cdot 10^{-9} cm$, as well as to the **relativistic** solution (at *β*=**0.007744**). The relativistic solution is characterized by the electron velocity $v_e/c = \alpha/\beta \approx 0.9407$ and the orbital radius $r = (c/v_e)^2 \sqrt{1-(v_e/c)^2} \cdot r_e \approx 10^{-13} cm$. Figure 2 shows the graphical solution of equations (7) and (11) using the coefficient $\beta(r)$ calculated using Eq. (10).

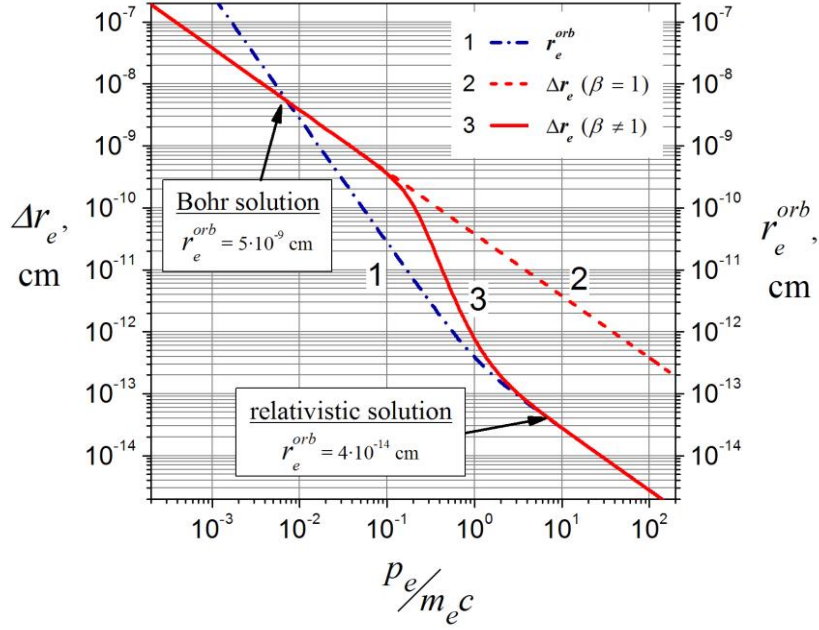

Figure 2. Dependencies of the uncertainty of the electron position $\Delta r_e$ and of the value of the electron orbit $r_e^{orb}$ on $p_e/m_e c$ calculated for $\beta = 1$ (i.e. neglecting the recoil effect) and at $\beta \neq 1$ (taking into account the recoil effect) using (7), (10) and (11). Curve 1 shows the radius of the electron orbit $r_e^{orb}$; curves 2 and 3 show the uncertainty of the electron position $\Delta r_e$ for $\beta = 1$ and $\beta \neq 1$, respectively.

The calculated parameters of the relativistic solution, as well as the estimates of the most characteristic parameters of the relativistic bound states of an electron and a proton are given in Table 1. In addition, in the right column of Table 1, we present the corresponding parameters from the International Committee on Data for Science and Technology (CODATA) [24], the experimental values of the corresponding parameters from [27, 28], which coincide with the calculations within the electroweak theory, as well as with the results of quantum chromodynamics calculations [29].

---

\*Ideas that a neutron can be interpreted as a bound state of a relativistic type, consisting of a proton and an electron, were expressed in the works of G.A. Gamow in the early 30s [30]. Estimated in [30], the electron velocity in such a bound state approximately corresponds to the value given in Table 1.

**Table 1. Parameters of relativistic solution of the problem of electron and proton bound states calculated using Eqs. (7), (10) and (11), which take into account the recoil effect.**

| Relativistic solution ($\beta = 0.007744$) | | | |
|---|---|---|---|
| Parameter | Formula | Calculation results | Results of calculations in quantum field theory and measurement data |
| 1. Electron velocity, orbital radius, orbital moments and energy parameters of the bound state of an electron with a proton | | | |
| Electron velocity | $v_e/c = \alpha/\beta = 0.0073/\beta$ | $v_e/c \approx 0.9407$ | |
| Orbital radius | $r = r_e(c/v_e)^2\sqrt{1-(v_e/c)^2}$, $r_e = e^2/m_e c^2$ | $r = 1.06 \cdot 10^{-13}\,cm$ | $0.84 \cdot 10^{-13}\,cm$ [24] |
| Binding energy $E_b = E_{kin} - E_{potential}$ | $E_b = m_e v_e^2 \left(2\sqrt{1-v_e^2/c^2}\right)^{-1}$ | $E_b = 1.3 m_e c^2$ (0.67 MeV) | Energy of electrons emitted in the beta decay of a neutron $0.1 \div 1.5 m_e c^2$ [27, 28] |
| Electron orbital angular momentum | $L_e = m_e v_e r \left(\sqrt{1-v_e^2/c^2}\right)^{-1} \approx \beta\hbar$ | $M_e = 0.0077 \cdot \hbar$ | |
| Electron magnetic moment | $M_e = L_e e/(2 m_e c)$ | $M_e = -4.77 \cdot \mu_N$ ($\mu_N$ is nuclear magneton) | |
| Sum of the magnetic moments of a proton, $M_p$, and an electron, $M_e$ | $M = M_p + M_e$ ($M_p = +2.81 \mu_N$) | $M = 2.81 - 4.77 \mu_N = -1.96 \mu_я$ | $-1.91 \mu_я$ [24, 29] |
| 2. Electron relativistic mass and the total mass of relativistic electron and proton | | | |
| Electron relativistic mass | $m_e^{rel} = m_e \left(\sqrt{1-v_e^2/c^2}\right)^{-1}$ | $m_e^{rel} = 2.9 \cdot m_e$ | |
| Total mass of the relativistic electron and proton in the bound state (in "quasi-atom") | $m = m_p + m_e^{rel}$ | $m = m_p + 2.9 \cdot m_e$ | $m = m_p + 2.5 \cdot m_e$ [24] |
| 3. Energy, inertial mass and sizes of virtual photons mediating a relativistic bound state of the electron and proton | | | |
| Virtual photon energy | $\varepsilon_{ph} = \hbar\omega_{ph} = E_{potential}/\beta$ | 340 $m_e c^2$ (0.17 GeV) | Virtual photons are close in their properties to particles. |
| Inertial mass of virtual photons | $m_{ph} \approx \hbar\omega_{ph}/c^2$ | $340 m_e$ | Masses of Yukawa interaction quanta 300-400 $m_e$ [27, 28] |

| Decay of Coulomb potential outside the quasi-atome | $\varphi(r) = \dfrac{e_1 e_2}{4r^2} e^{-r/r_c^{ph}}$ $r_C^{ph} = \hbar/(m_{ph} c) \approx 10^{-13}\,cm$ | Decay of the potential at $r > 10^{-13}\,cm$ | Decay of the Yukawa potential at $r > 10^{-13}\,cm$ [27,28] |
|---|---|---|---|
| 4. Energy released in decay of "quasi-atom" into a proton and an electron. | | | |
| Energy of electrons emitted from "relativistic quasi-atom" | $\Delta\varepsilon_e = (m_e^{рел} - m_e) c^2 =$ $= \dfrac{m_e c^2}{\sqrt{1-(\upsilon/c)^2}} - m_e c^2 =$ $= 2.9 m_e c^2 - 1 m_e c^2$ | $\Delta\varepsilon_e = 1.9 m_e c^2$ $(0 - 5.5\,МэВ)$ | Energy of electrons emitted in neutron beta decay $0.1 - 1.5 \cdot mc^2$ [27, 28] |

In the case of a nonrelativistic solution, the results of calculations based on Eqs. (7), (10) and (11) coincide with the Bohr parameters for the hydrogen atom. The data presented in Table 1 for the case of a relativistic solution are in a good agreement with the experimental data for the neutron, as well as with the known results of calculations of modern quantum field theory [24, 27-29].

Let us single out some parameters that reflect the correlation between the calculated parameters of the relativistic solution and known results in quantum field theory, which match with the observed data for neutrons. First of all, let us discuss the data related to the spatial distribution of positive and negative charged regions in a neutron. Results supporting the existence of positively- and negatively-charged regions in a neutron were obtained via the observation of elastic and inelastic scattering of ultrahigh-energy electrons by a neutron and proton [27, 31-36]. According to these data, the neutron has a negatively charged peripheral and a positively charged region outside the peripheral (i.e., it has analogies with the structure of ordinary atoms). Negative and positive charges cancel each other to produce a net zero charge, so that the free neutron is electrically neutral overall. In recent years, new results have been obtained on the spatial distribution of positive and negative charged regions in the neutron [37–38]. It has been discovered that the neutron has a negative charge both in its inner core and its outer edge, with a positive charge sandwiched in between to make the whole particle electrically neutral [37]. The possibility of the existence of a negative charge in a inner core of the neutron is now being discussed [39, 40]; it is not described within the framework of quantum electrodynamics and may be associated with peculiarities of the quark composition of the neutron in [37].

The following data, which seem to be interesting, concern the characteristic sizes of the relativistic bound state of a proton and an electron. These data, as in the case of the ground state in the Bohr model of the hydrogen atom, are characterized by the average value of the radius of electron rotation around the proton. According to our estimates based on equations (7), (10) and (11), the radius of the electron orbit is approximately equal to $10^{-13}\,cm$ that corresponds quite well to the neutron size of $\approx 0.84 \cdot 10^{-13}\,cm$, as stems from the calculations within the quark model as well as from the experimental results [24].

From the solutions of the equations (7), (10) and (11), it follows that the mass of the relativistic electron is $m_e^{rel} = 2.9 \cdot m_e$. The resulting total mass $m = m_p + 2.9 \cdot m_e$ of a proton and an electron (corresponding to the relativistic bound state of a proton and the electron) is in satisfactory agreement with the value of $m = m_p + 2.5 \cdot m_e$, which corresponds to QFT calculations and experimental data [24].

The magnetic orbital moment of the electron determined by the parameters of the relativistic solution is estimated to be $-4.77 \cdot \mu_N$ (where $\mu_N$ is nuclear magneton). Taking into account the magnetic moment of the proton $+2.81 \mu_N$, this leads to the resultant value of $-1.91 \mu_N$ for the magnetic moment of the relativistic bound state, which is consistent both in sign and in absolute value with the results of the quark theory for the magnetic moment of the neutron, $-1.91 \mu_N$, and agrees with the experiment [24, 29].

It seems important that, despite the approximate nature of the solution of the problem considered in this work, the obtained parameters of the relativistic solution correspond surprisingly well to the the known characteristics of neutrons, which can hardly be considered as accidental. It should be noted that the scales of the region of the relativistic solution given by our consideration and the size of the neutron as a particle, correspond to the boundary region between the characteristic scales of the electromagnetic, weak and strong fundamental interactions.

Thus, at sufficiently large distances, at which a proton and an electron can be considered as elementary particles, the electrodynamical description of the problem of formation of bound states of an electron and a proton within the virtual photon concept used here leads to the Bohr solution for a hydrogen atom. Within the same approximation (protons as structureless particles), the features of the formation of bound states of an electron and a proton at small subatomic distances are analyzed. At distances at which the contribution of higher-order terms in QED perturbation theory is negligible, and the main characteristics of the electron-neutron interaction are determined by the contribution of virtual photons, the most important factor being the recoil effects that appears at distances of the order of the electron Compton length. Accounting for these effects leads to a change in the uncertainty relations and decreases the values for the 3-momentum and 3-coordinate uncertainties of the electron at small distances. Under such conditions, the main parameters of the relativistic quantum bound state of an electron and a proton at distances $\sim 10^{-13} cm$ correspond with good accuracy to the parameters of a neutron. It can be expected that, despite somewhat contradictory aspects of the proton-electron model of a neutron, the estimates of the characteristic parameters of the relativistic solution obtained on the basis of the concept of virtual photons turn out to be satisfactory and, therefore, not very sensitive to imperfections of the proton-electron model for a neutron.

## 4. Conclusion

The main purpose of this work was to discuss the features of the formation of bound states in a system of particles, such as a proton and an electron, at atomic and subatomic distances. Our consideration was based on the QED concept of virtual photons that mediate electromagnetic interactions. The results obtained in this work can be formulated as follows.

It is shown that at atomic distances, the bound states of electrons and protons are formed by virtual photons with a pronounced maximum energy of the order of the potential energy of a particle at a given distance from the Coulomb center; the results of calculations of the bound state of an electron and a proton are consistent with the Bohr parameters in the hydrogen atom.

Within the framework of the virtual photon concept, the features of the uncertainty relations are analyzed also in the case of strongly interacting particles at subatomic distances. We found that at interparticle distances less than the electron Compton wavelength, the uncertainty relations become less constraining and allow for smaller values for the momentum and coordinate uncertainties, if compared to the usual Heisenberg uncertainty relations. This allows to describe the formation of bound states of particles (such as a proton and an electron) at small distances, at which the Coulomb potential is formed in the system and the main effects are due to virtual photons.

We confirm the existence of a relativistic bound state of an electron and a proton, which is characterized by a nuclear energy scale and distances $\sim 10^{-13} cm$ corresponding to a neutron. This conclusion is based on a reasonable agreement between the parameters of the relativistic solution found in this work and the measured characteristics of the neutron reported in the literature. Despite the approximate approach of our analysis of the problem of the interaction of proton and electron at small distances, the information obtained on their properties and peculiarities of the formation of bound states may be of interest in the context of physics of elementary particles, quantum mechanics and quantum electrodynamics.

We are grateful to A.A. Gorbatsevich, O. N. Krokhin, Yu.A. Kubyshin for the fruitfull discussions of the issues related to this work and helpful comments.